\documentclass[usenatbib]{mn2e}
\usepackage{natbibmnfix,graphicx,times}

\newcommand{\Mpc}{\,\mathrm{Mpc}}
\newcommand{\Msol}{\,\mathrm{M}_{\odot}}
\newcommand{\dG}{\,\bar{\delta}_g}
\newcommand{\rhobar}{\,\bar{\rho}_0}

\title[High-Redshift Voids]{High-Redshift Voids in the Excursion Set Formalism}

\author[D'Aloisio \& Furlanetto]{Anson
  D'Aloisio$^1$\thanks{Email: anson.daloisio@yale.edu} \& 
  Steven R. Furlanetto$^2$\\
$^1$Department of Physics, Yale University, PO Box 208120, New
    Haven, CT 06520-8120\\
$^2$Yale Center for Astronomy and Astrophysics, Yale University, PO Box 208121, New Haven, CT 06520-8121}

\begin{document}

\maketitle

\begin{abstract}
Voids are a dominant feature of the low-redshift galaxy distribution.  Several recent surveys have found
evidence for the existence of large-scale structure at high redshifts as well.
We present analytic estimates of galaxy void sizes at
redshifts $z \sim  5 - 10$ using the excursion set formalism.  We find that
recent narrow-band surveys at $z \sim 5 - 6.5$ should find voids with
characteristic scales of roughly $20$ comoving $\Mpc$ and maximum diameters
approaching $40 \Mpc$.  This is consistent with existing surveys, but a precise comparison is difficult because of the relatively small
volumes probed so far.  At $z \sim 7 -10$, we expect characteristic void
scales of $\sim 14 - 20$ comoving $\Mpc$ assuming that all galaxies within dark
matter haloes more massive than $10^{10} \Msol$ are observable.  We
find that these characteristic scales are similar to the sizes of empty regions
resulting from purely random fluctuations in the galaxy counts.  As a result,
true large-scale structure will be difficult to observe at $z \sim 7 -10$, unless galaxies in haloes with masses $\la 10^9 \Msol$ are visible.
Galaxy surveys must be deep and only the largest voids will provide meaningful
information.  Our model provides a convenient picture for estimating the
``worst-case'' effects of cosmic variance on high-redshift galaxy surveys with limited volumes.     
\end{abstract}


\section{INTRODUCTION}

The complex network of filaments and voids observed in the present-day
Universe is believed to have formed from an initially homogeneous distribution
of matter.  In hierarchical models of structure formation, tiny perturbations seeded by the inflationary epoch grew through
gravitational instability, collapsing first on smaller scales to form haloes.
The subsequent merging and clustering of smaller haloes resulted in the
formation of highly structured large-scale systems.  Perhaps the most striking
characteristic of the Universe today is the prevalence of large and nearly
spherical voids in the galaxy distribution.  The scales of these voids can be
enormous. Indeed, \citet{HandV2004} report characteristic radii of $R
\sim 15h^{-1}~\Mpc$ with maximum scales approaching $R \sim 25 h^{-1} \Mpc$ in
the 2dF Galaxy Redshift Survey.  

The characteristics of voids and the galaxies
that populate them have been the subject of numerous theoretical and
observational studies throughout the years \citep{gregory78, kirshner81,
  delapparent86, vogeley94, HandV2004, Conroy2005}.  To date, these studies have
mostly focused on low redshifts.  However, it is clear that voids should
appear at higher redshifts also.  The DEEP2 survey indicates that voids exist
at redshifts of $z \sim 1$ \citep{Conroy2005}.  Surprisingly, a handful of
recent Lyman $\alpha$ emitter (LAE) surveys have found hints of large-scale
structure at redshifts around $z \sim 5$ \citep{Shimasaku2003, Shimasaku2006, Hu2004, Ouchi2005}.

The modelling of voids poses an interesting theoretical problem. There have
been numerous studies utilising $N$-body simulations
\citep{MandW2002,Benson2003,Gottlober2003,Colberg2005}.  While these
simulations are invaluable tools for understanding the details of void dynamics, they are computationally expensive due to the large volumes and high dynamic range required to include a representative sample of voids while also resolving the much smaller galaxies that define them.

Analytic methods provide a useful alternative.  Perhaps the most promising
analytic model of void abundances is the excursion set approach taken by
\citet{SandV2004}.  They argue that voids actually provide deeper insight into
large-scale structure than halo formation itself.
Their assertion is based on the fact that underdense regions generally tend to
evolve toward a spherical geometry, making the idealisation of spherical
expansion more reasonable.  In contrast, gravitationally bound objects
typically have geometries that are far from spherical.  Approximating
gravitational collapse with the spherical model may be highly inaccurate,
which partially explains the discrepancies between the \citet{PandS1974} halo
mass function and simulations \citep{SandT1999,Sheth2001,Jenkins2001}.  

A key disadvantage to the approach of \citet{SandV2004} is the
difficulty in relating their definition of voids to observational
studies. As we will discuss in \S \ref{VoidDef}, \citet{SandV2004} use the
dark matter underdensity to define voids.  \citet{FandP2006} extend their model to define voids in terms of the local \emph{galaxy}
underdensity. They predict characteristic void sizes of $R \sim 10 h^{-1}
\Mpc$ at the present day -- nearly as large as observed voids.  

In what follows, we present analytic estimates of void size
distributions at redshifts between $z \sim 5 - 10$.  Our aim is to provide
a convenient basis of comparison for current and future high-redshift
observations -- presumably, though not limited to, LAE surveys.  As such, we
consider the effects of statistical fluctuations in the galaxy counts and the
abundance of Ly $\alpha$ emitting galaxies on void observations.  

LAE surveys have become an invaluable tool in cosmological studies.  In
addition to building larger samples at $z \sim 5$, observers have pushed the
threshold to redshifts as high as $z \sim 7 - 10$
\citep{WandC2005,Cuby2007,Stark2007,Ota2007}.  Surveys at these redshifts
could potentially reveal important information on the epoch of reionization.
Indeed, the observed clustering properties of LAEs could someday be a powerful
probe of the epoch \citep{furl04-lya, Furlanetto2006,McQuinn2006,McQuinn2007, mesinger07}.
Regions of ionized hydrogen grow quickly around clustered galaxies as reionization progresses.  When these regions are large enough, Ly $\alpha$
photons are sufficiently redshifted before they reach neutral hydrogen gas,
allowing them to avoid absorption in the intergalactic medium (IGM).  Sources
within overdense regions are therefore more likely to be observed relative to
void galaxies, resulting in a large-scale modulation of the number density.  One method to quantify such clustering is with void statistics, as first attempted by \citet{McQuinn2007}.  This provides comparable power to correlation function measurements of the galaxies.  However, taking full advantage of this technique requires a deeper understanding of voids in the underlying galaxy distribution; our calculations aim to provide such a baseline model.

The remainder of this paper is organized in the following manner.  In \S
\ref{ExcursionSet}, we briefly  review the basic principles of the excursion
set formalism.  In section \ref{VoidDef}, we present the \citet{FandP2006}
definition of voids in terms of the local galaxy underdensity.  Section
\ref{VoidDistributions} contains the main results of this paper: void size
distributions at $z = 4.86 - 10$.  In \S \ref{StochasticVoids}, we estimate
the typical sizes of voids that result from random fluctuations in the galaxy
distribution and develop an alternative definition of voids.    In \S
\ref{VisibleFrac}, we explore the assumption that only a certain fraction of
galaxies are actually visible in LAE surveys.  Section
\ref{Observations} contains a rough comparison of our calculations to high
redshift Ly $\alpha$ surveys.  Finally, we offer concluding remarks in \S \ref{Discussion}.  

In what follows, we assume a cosmology with parameters $\Omega_m
= 0.24,~\Omega_{\Lambda} = 0.76,~\Omega_b = 0.042,~H =
100h~\mathrm{km~s}^{-1}~\Mpc^{-1}$ (with $h = 0.73$), $n = 0.96$, and
$\sigma_8 = 0.8$, consistent with the latest measurements \citep{Spergel2007}.
All distances are reported in comoving units.    

\section{VOIDS AT HIGH REDSHIFTS}

\subsection{Voids in the excursion set formalism}
\label{ExcursionSet}

%%%%%%%%%%%%%%%%TABLE 1
\begin{table*}
\caption{  Comoving LAE Densities and Mass Thresholds at High
    Redshifts}
\begin{tabular}{@{}lccc}
\hline
$z$ & $n_g(PS; 10^{-4} \Mpc^{-3})$ 
&  $n_g(ST; 10^{-4} \Mpc^{-3})$ & $m_{min}(\times 10^{10}~\Msol)$\\
\hline
4.86 & 1.0 &  & 32 \\
5.7 & 5.4 &  & 7.9 \\
6.5 & 2.6 &  & 6.5 \\
\hline
Fixed $m_{min}$\\
\hline
7 & 61 & 108 & 1.0 \\
8 & 19 & 47 & 1.0 \\
9 & 5.4 & 19 & 1.0 \\
\hline
Fixed $z$\\
\hline
10 & 10,413 & 14,709 & 0.01 \\
10 & 220 & 495 & 0.1\\
10 & 1.3 & 6.8 & 1.0 \\
10 & $6.3 \times 10^{-4}$ & $1.6 \times 10^{-2}$ & 10\\
\hline
\end{tabular}

\medskip
Where applicable, columns 2 and 3 show the number densities obtained with the Press-Schechter and Sheth-Tormen mass functions respectively.
\label{MminTable}
\end{table*}
%%%%%%%%%%%%%%%%%%%%%%%%%%%%%%%%

In this section, we briefly summarise the extension of excursion set
principles to voids pioneered by \citet{SandV2004}.  For an
excellent review on the excursion set formalism and its many applications, we
refer the reader to \citet{Zentner2007}.   

The approach we describe here is in many ways similar to the excursion set formulation of the dark matter
halo mass function \citep{PandS1974, Bond1991}.  At a fixed point in space,
the linear density contrast $\delta^L$ is smoothed on a scale $R$.  We will denote the smoothed version of the linear density contrast
with $\delta^L(R)$.  The variance of the smoothed density contrast
$\sigma(R)^2$ is simultaneously computed for each scale $R$.  The set of
points $\left[\delta^L(R), \sigma(R)^2\right]$  define a trajectory
parametrized by $R$ in the $(\delta^L, \sigma^2)$ plane.

It is often convenient to work in coordinates in which the density contrast
$\delta^L$ is linearly extrapolated to the present day.  The linear density
contrast at some redshift $z$ is related to the extrapolated version through $\delta^L(z) = D(z) \delta^L_0$, where $D(z)$ is the
growth factor from linear perturbation theory normalised to unity at $z = 0$ and $\delta^L_0$ is the linear
density contrast extrapolated to present day.  In this paper, we will work almost exclusively in
linearly extrapolated coordinates.  For brevity, we will drop the
subscript ``0''.  Whenever it is necessary to consider quantities that have
not been linearly extrapolated, we will note it in the text.

We define a void within the excursion set formalism to be a region of
scale $R$ with smoothed density contrast $\delta^L(R)$ that has fallen below a
potentially scale-dependent density contrast threshold $\delta^L_v$, henceforth referred to
as the void barrier.  This barrier is analogous to the critical overdensity
$\delta_c$ used in the \citet{PandS1974} formalism.  To proceed, we must
therefore specify the analogous threshold for voids.  There are two sensible choices.  The first option is to
consider the physics of dark matter.   Alternatively, one may
rely upon observational parameters, as we will in \S \ref{VoidDef}.

In order to avoid counting smaller voids that are embedded within larger scale
voids -- the so-called void-in-void problem -- we only consider the largest scale
(smallest $\sigma^2$) at which a given trajectory crosses $\delta^L_v$.
Hence, the most important step in calculating the void
mass function with excursion set techniques is to obtain the distribution of
scales $R$ at first crossing.  

There is, however, an important difference between the halo and void
formalisms.  In deriving the void mass function, we must be careful to exclude voids that are embedded in larger overdense regions that will eventually
collapse into virialized objects.   These voids will be crushed out
of existence during the collapse of larger scale overdensities (this is the
so-called void-in-cloud problem).  To address this,
\citet{SandV2004} introduced a second absorbing barrier, henceforth referred to as the void-crushing barrier
$\delta^L_p$.  They argue that an appropriate choice for the
void-crushing barrier lies between $\delta^L_p = 1.06 - 1.69$.  We therefore
seek the distribution of scales $R$ at which trajectories first cross
$\delta^L_v$ without having crossed $\delta^L_p$.  The problem of calculating
the void mass function is reduced to solving the diffusion equation in the
$(\delta^L, \sigma^2)$ plane with appropriate boundary conditions at two absorbing barriers.  

\subsection{Defining voids in terms of the galaxy underdensity}
\label{VoidDef}

In this section, we obtain the linear underdensities defining voids in
the excursion set formalism. \citet{SandV2004} use the scale independent underdensity corresponding to shell-crossing in the spherical expansion
model ($\delta^L_v = - 2.81$).  Shell-crossing occurs when fast moving
mass shells from the interior of voids run into initially larger and slow
moving mass shells, forming an overdense ridge.  While it is certainly
reasonable to define voids through shell crossing, analytic
calculations utilising this criteria yield characteristic void sizes at the
present ($\sim 3 h^{-1} \Mpc$) that are significantly smaller than observed
($R \sim 15 h^{-1} \Mpc$). Moreover, a void
formation criterion based on the evolution of dark matter underdensities is
difficult to reconcile with observational data.  Of course, observational
surveys are only sensitive to \emph{galaxy} underdensities.  Following this
reasoning, \citet{FandP2006} extend the analytic model of \citet{SandV2004}
by utilising a void formation criteria that defines voids in terms of galaxy
underdensities.

Consider a region of space with a linear density contrast $\delta^L(R)$.  The mass contained within the region is given by $M
= \rhobar(1 + \delta) (4 \pi/3) R^3$, where $\rhobar$ is the average
matter density today and $\delta$ is the \emph{true} density contrast.  We use the
spherical expansion model in order to relate the true density contrast in a
region to the local linear density contrast $\delta^L$ (pre-extrapolation).
In this paper, we use a fit to the function $\delta\left[\delta^L \right]$
given by equation (18) of \citet{MandW1996}.  Although they assume an
Einstein-de Sitter cosmology, the fit is an excellent approximation to the
$\Lambda$CDM version at high redshifts (accurate to a few percent).  The
spherical model and fit both break down at $\delta^L = -2.81$, corresponding
to shell-crossing, since mass is no longer conserved (the underdensities
relevant to this work never reach such low values).  The function
$\delta\left[\delta^L \right]$ allows us to write the mass contained within a spherical region as  $M = \rhobar~(4 \pi/3) (R/\eta)^3$, where $\eta(\delta^L) = \left[1 +\delta(\delta^L)\right]^{-1/3}$.  

Using the halo model of structure formation, the comoving number density of observable galaxies within the considered region is \citep{FandP2006}
\begin{equation}
n_g(m_{min}|\delta^L, M) =
\int^{\infty}_{m_{min}}{dm_h~\left<N(m_h)\right>~n_h(m_h|\delta^L, M)}
\label{CumGalDens}
\end{equation}
where $m_h$ is the dark matter halo mass, $\left<N(m_h)\right>$ is the average
number of galaxies per halo, $n_h$ is the conditional halo mass
function \citep{Bond1991, LandC1993}, and $m_{min}$ is a halo mass detection threshold.  For the
high redshifts of interest here, we assume that the average number of
observable galaxies per halo is unity above $m_{min}$ and zero below it. 

Where possible, we will fix $m_{min}$ by normalising the comoving galaxy density to observational data.  Recently, there have been several
claimed identifications of large-scale structure in LAE surveys.  For redshifts of $z = 4.86, 5.7$, and $6.5$, we will use the comoving
number densities from the surveys of \citet{Shimasaku2004}, \citet{Shimasaku2006},
and \citet{Kashikawa2006} respectively.  Both of the first two surveys found evidence for large-scale structure at high redshifts.  

For $z = 5.7$, we integrate the Schechter
function fit to the luminosity function obtained by \citet{Shimasaku2006} to $L = 3\times10^{42}$ ergs/s. Similarly, \citet{Kashikawa2006}
provide upper and lower limits for the luminosity function at $z =
6.5$.  We use the upper estimate with a fixed $\alpha = -1.5$.  Owing to the
lack of observational data at higher redshifts, we do not normalise the galaxy
number densities to observational data.  Instead, we simply specify various
halo mass thresholds to define a set of mythical high-redshift surveys.  Table \ref{MminTable} shows the comoving number densities and halo mass thresholds obtained for redshifts between $z = 4.86$ and $z = 10$. 

For the conditional halo mass function in equation (\ref{CumGalDens}), we use
the excursion set expression \citep{Bond1991, LandC1993}
\begin{eqnarray}
n_h(m_h|\delta^L, M) &  =  & \sqrt{\frac{2}{\pi}}~\frac{\rhobar}{m^2_h} 
\left|\frac{d\ln \sigma}{d\ln m_h} \right| \frac{\sigma^2(\delta^L_c -
    \delta^L)}{\left[\sigma^2 - \sigma(M)^2\right]^{3/2}} \nonumber \\
  & & \times \exp\left[-\frac{(\delta^L_c - \delta^L)^2}{2[\sigma^2 -
      \sigma(M)^2]}\right],
\label{CondMassFunc}
\end{eqnarray}                        
where $\delta^L_c$ is the critical density contrast at collapse, extrapolated
to the present day.  Although equation (\ref{CondMassFunc})
provides a reasonable approximation to the halo abundance, it is certainly not
the most accurate choice.  \citet{SandT1999} and \citet{Jenkins2001} obtain
more accurate fits to the results of numerical simulations (see \citealt{CandW2007} for recent tests at high redshifts).  Table \ref{MminTable} compares the
mean number densities obtained from the different mass functions.  Owing to a
larger high-mass tail with respect to the Press-Schechter form, the
Sheth-Tormen mass function clearly results in larger number densities.  However, as \citet{FandP2006} point
out, normalising the comoving galaxy number density to observed values significantly decreases the differences between mass functions.
Our choice of analytic mass function in equation (\ref{CondMassFunc})
therefore suffices for the calculations in this paper.

We can now use equation (\ref{CumGalDens}) to write down a relationship
between the observable mean galaxy underdensity $\dG$ in a region of size $R$ and the linearised dark matter density contrast $\delta^L(R)$ \citep{FandP2006}:
\begin{equation}
1 + \dG(m_{min}, \delta^L, R) = \frac{n_g(m_{min}|\delta^L,
  M)}{\eta^3~n_g(m_{min})}.
\label{deltaL}
\end{equation}
Note that the factor of $\eta^{-3}$ is used to transform the numerator on the right hand side from Lagrangian to Eulerian coordinates.  

Computation of the void barrier requires a suitable choice for the galaxy
underdensity defining a void.  Ideally, $\dG$
would be chosen to most accurately reflect the void finding algorithm in the
survey of interest.  Unfortunately, the aforementioned LAE surveys
are not large enough to allow a systematic search for voids.   Hence, an
appropriate choice for $\dG$ is not at all clear. Following
\citet{FandP2006}, we choose $\dG = -0.8$ as the fiducial value for
calculations in this paper; we will consider a modified definition in \S
\ref{EmptyFrac}.

\begin{figure} 
\begin{center}
\resizebox{8.5cm}{!}{\includegraphics{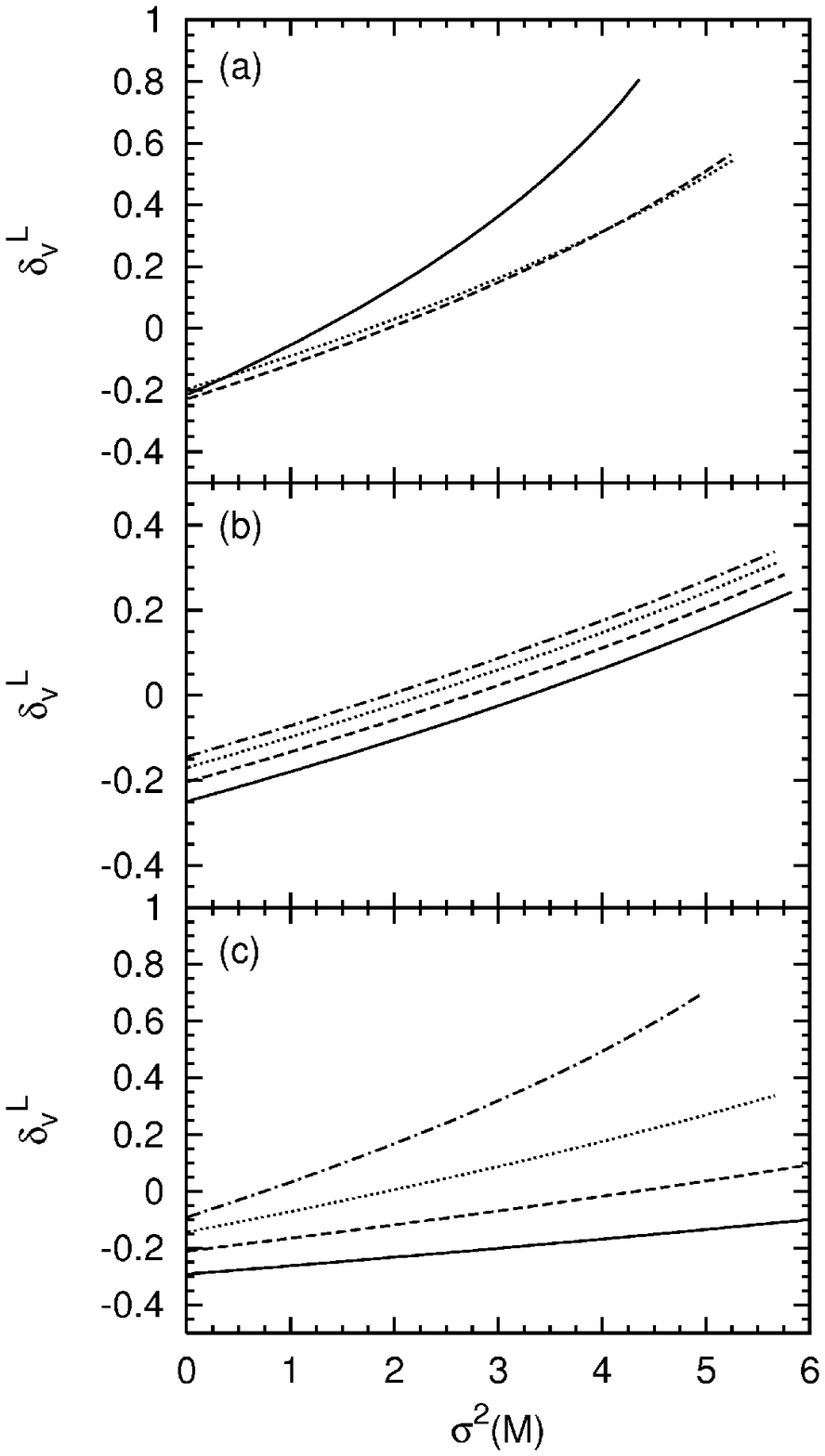}}
\end{center}
\caption{Linear underdensity thresholds (pre-extrapolation) defining voids in the excursion set
  formalism.  All curves assume a galaxy underdensity of $\dG = -0.8$.  $(a)$:
  The solid, dashed, and dotted curves show void barriers at $z = 4.86, 5.7$,
  and $6.5$ respectively.  Comoving galaxy number densities are normalised to
  data in the surveys of \citet{Shimasaku2004}, \citet{Shimasaku2006}, and
  \citet{Kashikawa2006}.  $(b)$:  Void barriers at redshifts of $z = 7, 8, 9$, and $10$
  (solid, dashed, dotted, dot-dashed respectively).  A fixed halo mass
  threshold of $m_{min} = 10^{10} \Msol$ is assumed.  $(c)$:  The solid,
  dashed, dotted, and dot-dashed curves show void barriers at a fixed redshift
  of $z = 10$ for halo mass thresholds of $m_{min} =  10^8, 10^9, 10^{10}$ and
  $10^{11}~\Msol$ respectively.}
\label{VoidBarriers}
\end{figure}

The prescription for defining voids in terms of the galaxy underdensity is now
straightforward.  We first set $n_g(m_{min})$ equal to the mean galaxy density
extracted from observational surveys and solve for the corresponding
$m_{min}$.   We then define a void to be a region with a given galaxy
underdensity $\dG$ and solve equation (\ref{deltaL}) for the corresponding
$\delta^L_v(R)$ to be used in the excursion set
formalism \citep{FandP2006}.  Several examples of such calculations are shown
in Figure \ref{VoidBarriers}. Figure \ref{VoidBarriers} $(a)$ shows
void barriers at $z = 4.86$, $5.7$ and $6.5$.  Panel $(b)$ shows higher redshift void barriers at $z =  7 - 10$ for a fixed halo mass
threshold of $m_{min} = 10^{10} \Msol$. Panel $(c)$  shows the void barriers
for several halo mass thresholds and a fixed redshift of $z = 10$.  For
reference, $\sigma^2 \approx 1.58, 0.72,$ and $0.42$ for $R = 5, 10,$ and
$15 \Mpc$ respectively.  Owing to the increased bias of galaxy haloes relative to the
underlying matter density at high redshifts, the matter density contrast
required to produce a mean galaxy underdensity of $\dG = - 0.8$ is smaller compared to the $z = 0$ case \citep[see Figure 2 in][]
{FandP2006}. 

%%%%%%%%%%%%%%%%% FIGURE 1
%%%%%%%%%%%%%%%%%%%%%%%%%%%

Figure \ref{VoidBarriers} illustrates that for large $R$, voids must be
underdense in dark matter as one would expect. However, we note that in the
formalism of \citet{FandP2006}, small voids may actually correspond to regions
that are overdense in dark matter ($\delta^L_v > 0$).  This is due to finite
size effects. For both cases what is important is that voids are defined to be
regions that are underdense in \emph{galaxies} ($\dG < 0$).  We shall see
that, for most scales of interest, voids in the galaxy distribution do in fact correspond to dark matter underdensities.       
 
\subsection{Void size distributions}
\label{VoidDistributions}

Using the dark matter underdensities obtained in \S \ref{VoidDef} as void
barriers, we are now in the position to calculate void size distributions within the excursion set formalism.  Most of the void
barriers shown in Figure \ref{VoidBarriers} are well approximated as linear
functions of $\sigma^2(M)$.  One approach is to solve a diffusion
problem in the $(\delta^L, \sigma^2)$ plane with one linear absorbing barrier -- the void barrier -- and one
constant absorbing barrier -- the void-crushing barrier.  Owing to the
non-trivial boundary conditions, obtaining an analytic solution for this problem is rather
difficult.  Numerical techniques for obtaining the first-crossing distribution with generic boundary conditions do exist \citep[for an overview
of such techniques, see][]{Zentner2007}.  However, in the interest of
obtaining analytic solutions, we approximate the first-crossing distribution
with a solution involving two linear absorbing barriers of the form
$\delta^L_v = A_v + \beta \sigma^2$ and $\delta^L_p = A_p + \beta \sigma^2$.  We find
that for $A_p = 1.06$, $\delta^L_p < 1.69$ over the range of interest for the
models considered in this paper.  More importantly, we show in \S
\ref{VoidCrushing} that the void-crushing barrier has little effect on the
calculated void size distributions anyway.

A full derivation of the mass function in the case with two linear absorbing
barriers with identical slopes can be found in \citet{FandP2006}.  We provide a brief summary here.  In the following discussion, we set $S \equiv \sigma^2$ for simplicity.

The probability that a trajectory will cross the void barrier first (i.e. before the void-crushing barrier) at a scale between $S$ and $S + dS$ is given by \citep{FandP2006}
\begin{eqnarray}
F_v(S)\,dS & = & \sum^{\infty}_{n = 1}{\frac{n^2 \pi^2 D^2}{A_v^2} \frac{\sin(n \pi
    D)}{n \pi} \exp\left[ - \frac{n^2 \pi^2 D^2}{2 A_v^2 / S}\right]} \nonumber
  \\ & & \times \exp\left[- \beta A_v - \beta^2 S / 2 \right]~dS.
\label{FirstCrossing}
\end{eqnarray}
Here, $D \equiv \left|A_v\right|/(A_p + \left|A_v\right|)$ and $F_v(S)$ is known as the first-crossing distribution.

Following equation (\ref{FirstCrossing}), the desired mass function has the form \citep{FandP2006}
\begin{equation}
n_v(M) = n^{2CB}_v(M) \exp\left( - \beta A_v - \beta^2 \sigma^2 / 2
\right),
\label{LBMassFunc}
\end{equation}
where the function $n^{2CB}_v(M)$ is given by 
\begin{eqnarray}
n^{2CB}_v(M)& = & \frac{2\,\rhobar}{M^2}~\left| \frac{d\ln \sigma}{d\ln M} \right|
\sum^{\infty}_{n = 1}{\frac{n^2 \pi^2 D^2}{(A_v / \sigma)^2}
  \frac{\sin(n \pi D)}{n \pi}} \nonumber \\
 & & \times \exp\left[ - \frac{n^2 \pi^2 D^2}{2(A_v /
      \sigma)^2} \right].
\label{CBMassFunc}
\end{eqnarray}
Note that equation (\ref{CBMassFunc}) is the mass function obtained in the
case with two constant barriers $\delta^L_v = A_v$ and $\delta^L_p = A_p$ \citep{SandV2004}.

%%%%%%%%%%%%%%%%%% FIGURE 2
\begin{figure}
\begin{center}
\resizebox{9.0cm}{!}{\includegraphics{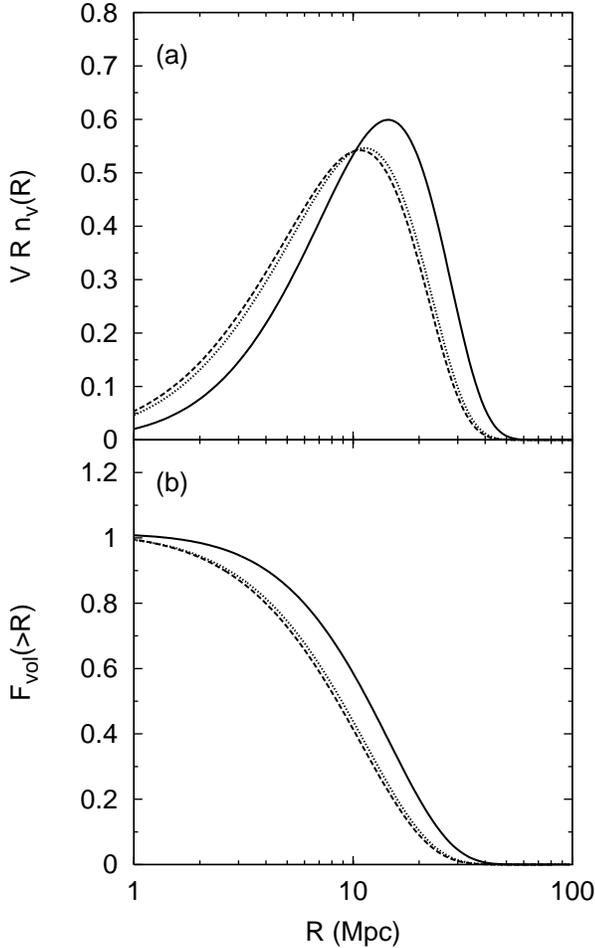}}
\end{center}
\caption{Void size distributions and volume filling fractions at redshifts of $z = 4.86$ (solid), 5.7 (dashed), and 6.5 (dotted).  See Table \ref{MminTable}
  for model parameters.}
\label{OBS}
\end{figure}
%%%%%%%%%%%%%%%%%%%%%%%%%%

\begin{figure*}
\begin{center}
\resizebox{17cm}{!}{\includegraphics{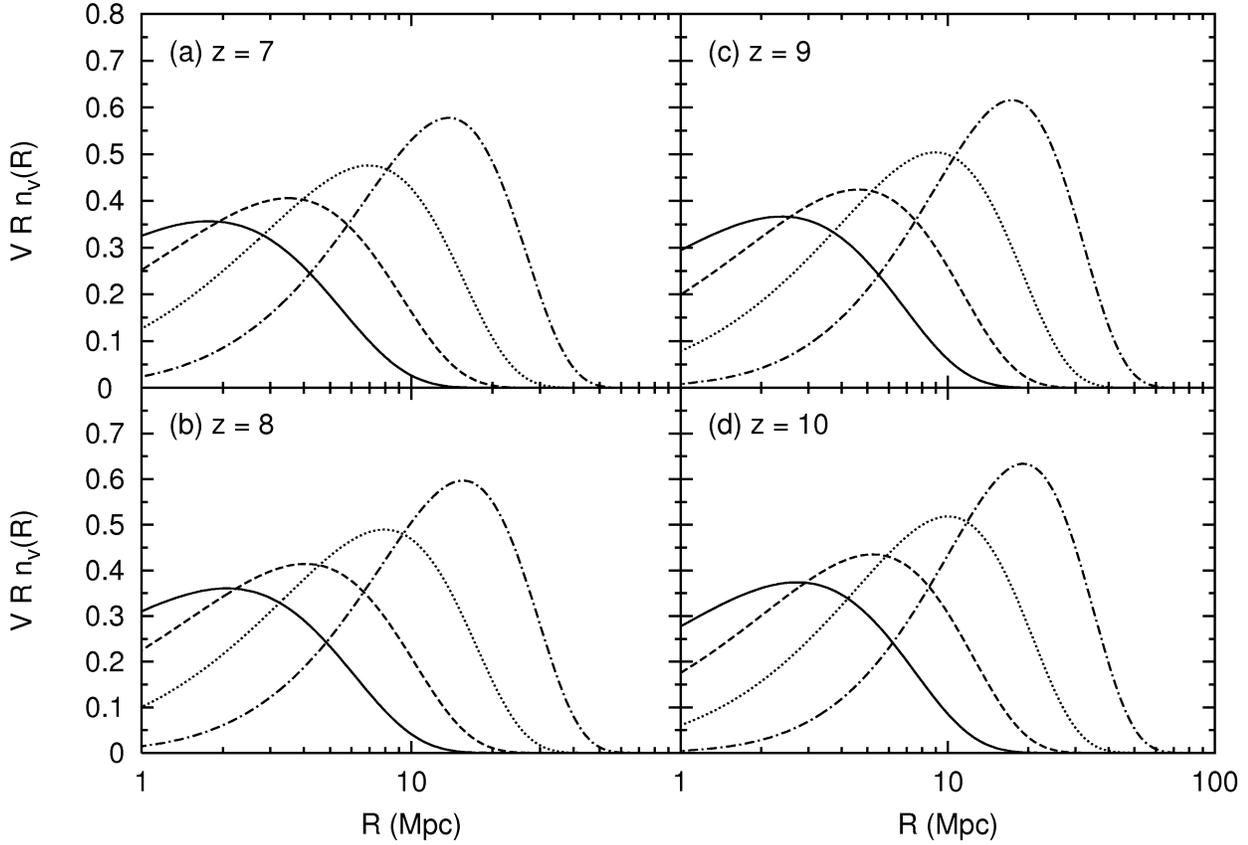}}
\end{center}
\caption{$(a) - (d)$: Void size distributions at redshifts of $z = 7 - 10$ for
  several different halo mass thresholds.  Solid, dashed, dotted, and
  dot-dashed curves assume $m_{min} = 10^8, 10^9, 10^{10}$ and $10^{11} \Msol$
  respectively.}
\label{Z7to10}
\end{figure*}

Converting equation (\ref{FirstCrossing}) to units of distance to
obtain the fraction of voids per logarithmic interval in $R$ yields $V
R~n_v(R)$, where the volume $V$ and radius $R$ are in comoving coordinates. 
Hence, the fraction of volume contained within voids of radius greater than $R$
is given by
\begin{equation}
F_{vol}(> R) = \int_R^{\infty}{ V n_v(r)~dr} = \int_M^{\infty}{
  \frac{m}{\rhobar} \eta^3 n_v(m)~dm}.
\label{VolumeFilling}
\end{equation}
In Figure \ref{OBS}, we plot $VR~n_v(R)$ and $F_{vol}(>R)$ for $z = 4.86, 5.7$,
and $6.5$ using the parameters given in Table \ref{MminTable}. Typical voids
in the LAE distribution at $z \sim 5$ are roughly $20$ comoving
$\Mpc$ across in our calculations.  The $z = 4.86$ voids are largest because
the survey of \citet{Shimasaku2004} has the highest detection threshold (see Table
\ref{MminTable}). Similarly, Figure \ref{Z7to10} shows $VR~n_v(R)$ for $z = 7, 8, 9$ and $10$ for a
variety of halo mass thresholds.   Using
equation (\ref{VolumeFilling}) and $m_{min} = 10^{10} \Msol$, we find that approximately 21, 26, 31 and 37 \% of
space is filled by voids with radii larger than 10 Mpc at $z = 7, 8, 9$, and
$10$ respectively.   Panel $(b)$ in Figure \ref{OBS} shows that $F_{vol}(>R)$ approaches unity as
$R \rightarrow 0$, indicating that the entire universe is filled by voids.  As
\citet{FandP2006} point out, this peculiarity is primarily due to the fact
that we have included voids embedded within regions that are not quite at
turnaround.  These voids will likely be suppressed by the surrounding
overdensities.  By allowing them to expand fully, we have overestimated the
volume contained within voids.   

%%%%%%%%%%%%%%%% FIGURE 3
%%%%%%%%%%%%%%%%%%%%%%%%%%%%

The peaks in Figures \ref{OBS} and \ref{Z7to10} occur where $\sigma \sim \delta^L_v$. The large scale cutoffs are due to the
smoothness of the matter density field at large scales.  As $R \rightarrow
\infty$, $\sigma^2(R) \rightarrow 0$ and the probability of crossing the
void barrier approaches zero.  On the other hand, the small-scale cutoffs are a result of the rising void barrier.  Note that
this differs from the low-$z$ results of \citet{SandV2004} and \citet{FandP2006}, in
which the small-scale cutoffs are due to the void-crushing barrier. 
As we have seen in \S \ref{VoidDef}, high-redshift galaxy voids are not
as underdense in dark matter as their present day counterparts and the void
barrier actually crosses through $\delta^L_v = 0$.  Most trajectories are
therefore absorbed by the void barrier before reaching larger $\sigma^2(R)$, resulting in a
suppression of the mass function for small scales.  Section \ref{VoidCrushing} of this paper examines the role of the void-crushing barrier in detail.

%%%%%%%%%%%%%% FIGURE 4
\begin{figure}
\begin{center}
\resizebox{8.5cm}{!}{\includegraphics{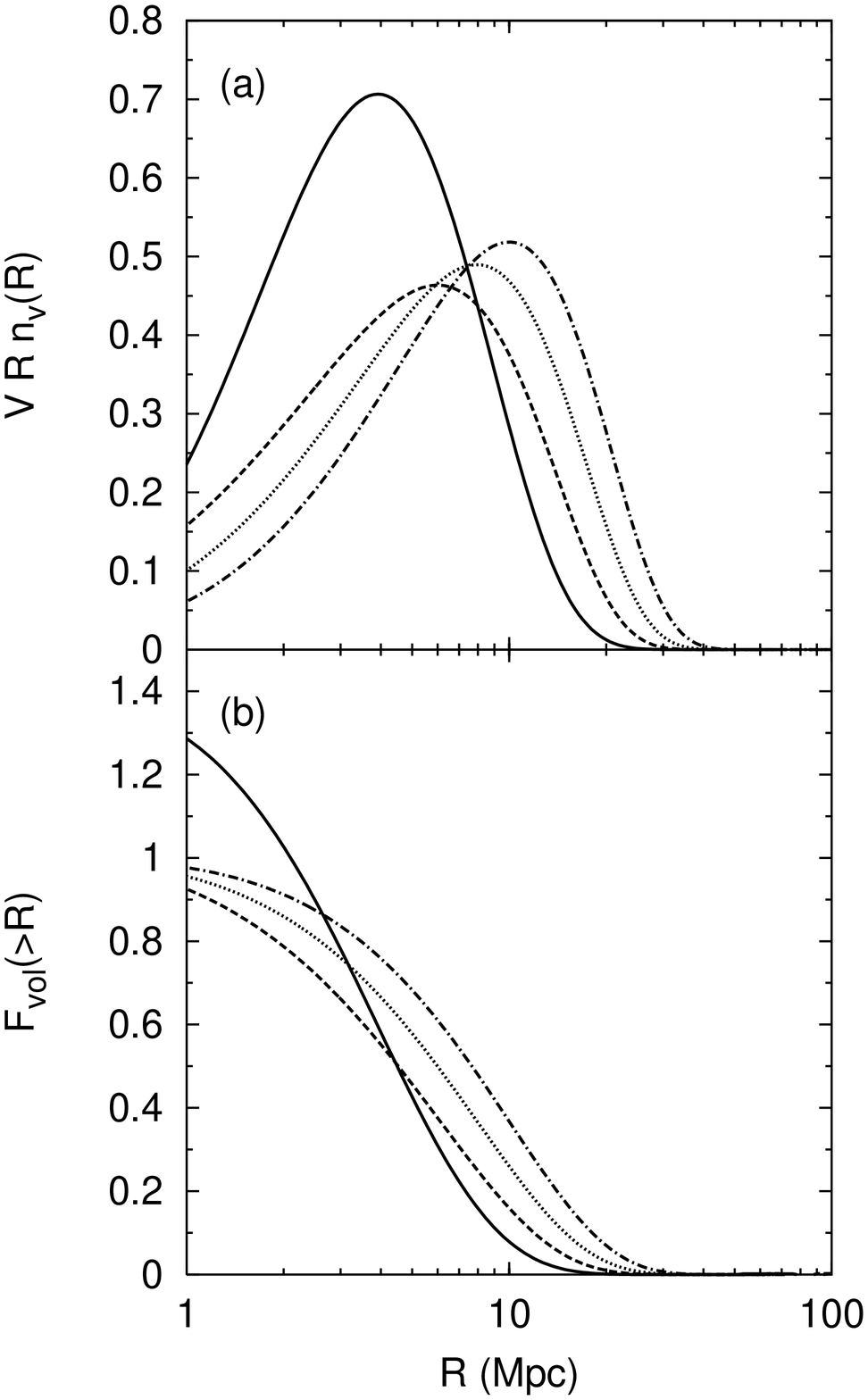}}
\end{center}
\caption{Void size distributions and volume filling fractions at $z = 0$
  (solid), $6$ (dashed), $8$ (dotted), and $10$ (dot-dashed) for a fixed
  $m_{min} = 10^{10} \Msol$.  Voids are larger at high redshifts due to a
  decreased galaxy number density and increased bias with respect to the
  underlying dark matter.}
\label{ZEvolution}
\end{figure}
%%%%%%%%%%%%%%%%%%%%%

Figure \ref{ZEvolution} depicts the evolution of the void size
distributions with redshift for a fixed halo mass threshold of $m_{min} =
10^{10} \Msol$.  The solid, dashed, dotted, and dot-dashed curves correspond
to $z = 0$, $6$, $8$, and $10$ respectively.  Panel $(a)$ shows that the
characteristic scales of voids are actually slightly larger at higher
redshifts (decreasing by roughly $2 \Mpc$ per $\Delta z = - 2$ between $z = 10$ and $z =
6$) due to a decreasing galaxy number density and increasing
bias.\footnote{We found an error in the $z$-dependent mass
  function used by \citet{FandP2006}.  As a result, their Figure 7 incorrectly indicates that galaxy
  void sizes decrease with redshift.}.   Interestingly, the void scales decrease by only $2 \Mpc$ between $z = 6$ and $z = 0$.  This is due to the competition between a decreasing
spatial bias of galaxies with respect to the underlying matter and the gravitational expansion of underdense regions.
Neglecting gravitational effects, the
increased abundance and decreased spatial bias of galaxies at lower redshifts
would decrease the characteristic scales of voids.  However, as we
approach the present day, underdensities evolve through gravitational
expansion; voids become deeper and larger as mass is evacuated from the
interior.  These two effects work against each other, creating less net change
in the characteristic scale of voids between $z = 6$ and $z = 0$.

\subsection{The void-crushing barrier} 
\label{VoidCrushing}

In equation (\ref{LBMassFunc}), we assume that both the void and void-crushing
barrier are linear functions of $\sigma^2$ with the same slope.  In this
section we test how our results depend on the particular choice of
void-crushing barrier.  In the following discussion, all calculations will be performed using our fiducial void underdensity of $\dG = -0.8$ and at a redshift of $z = 5.7$.  See Table \ref{MminTable} for the model parameters.

First, recall that we have used a linear function for the void-crushing barrier, introducing some scale dependence to the void-crushing with little physical basis.  Fortunately, the relevant range of $\sigma^2$ is small enough at high redshift that this makes little difference:  with $A_p = 1.06$, we have $\delta_p = 1.15$ at $R = 10 \Mpc$ for $z=5.7$.

\begin{figure}
\begin{center}
\resizebox{8.5cm}{!}{\includegraphics{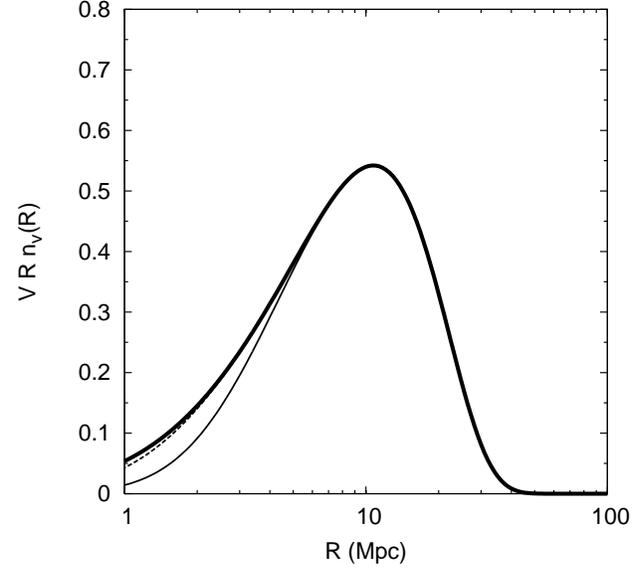}}
\end{center}
\caption{Void size distributions at $z = 5.7$ for different void-crushing
  barriers of the form $\delta^L_p = A_p + \beta S$.  The thin solid, dashed,
  dotted, and dot-dashed curves assume $A_p =$ 0.3, 0.5, 1.06 and 1.69
  respectively. The thick solid curve shows the void size
  distribution obtained by neglecting the void-crushing barrier entirely.
  Note that the dotted and dot-dashed curves are obscured by the thick curve,
  indicating that the size distribution is essentially independent of the
  void-crushing barrier for reasonable choices of $A_p$.}
\label{Ap}
\end{figure}

We have previously used $A_p = 1.06$, corresponding to the linear density
contrast at turnaround in the spherical model.  Our first task is to vary this
value.    The thin solid, dashed, dotted, and dot-dashed curves in Figure
\ref{Ap} show the void size distribution at $z = 5.7$ for $A_p = 0.3$, $0.5$,
$1.06$, and $1.69$ respectively. Note that the curves with $A_p = 1.06$ and
$1.69$ are identical and lie within the thick solid curve.

Figure \ref{Ap} shows that the particular choice of $A_p$ has a minor
effect on the void size distribution at most.  If $A_p$ is comparable to
$|A_v|$, a small number of trajectories will encounter the void-crushing
barrier before the void barrier.  These trajectories are subtracted from
$n_v(R)$ since they represent voids that will eventually be crushed out of
existence.  In Figure \ref{Ap}, this suppression is visible at small $R$ for $A_p = 0.3$ and $A_p = 0.5$.  We note that no suppression is seen in
the cases where $A_p = 1.06$ and $A_p = 1.69$.  We will see in \S
\ref{PoissonVoids} that the issue of small-scale suppression is
irrelevant anyway because of stochastic fluctuations in the galaxy distribution.  

On the other hand, if $A_p \gg |A_v|$, then the probability that a trajectory
will encounter the void-crushing barrier first at small $R$ is negligible.  In
this case, the small scale cut-off of the void size distribution is not due to
the void-crushing barrier.  As $R$ decreases, $\delta^L_v$ crosses zero,
trapping most trajectories where $\delta^L_v \sim 0$.  Hence, the void barrier
itself absorbs most trajectories before they reach small $R$.
  
%%%%%%%%%%%%% FIGURE 5
%%%%%%%%%%%%%%%%%%

Following the argument above we would expect that the void size
distribution becomes independent of the void-crushing barrier as $A_p$ gets larger.  To illustrate that this is indeed the case, it is
instructive to consider the diffusion problem with only one linear absorbing barrier of the
form $\delta^L_v = A_v + \beta S$.  The appropriate mass function is \citep{Sheth1998}
\begin{equation}
n^{1LB}_v(M) = \sqrt{\frac{2}{\pi}} \frac{\rhobar~|A_v|}{M^2 \sigma} ~\left|
    \frac{d\ln \sigma}{d\ln M} \right|~\exp\left[- \frac{(\beta \sigma^2 +
    A_v)^2}{2 \sigma^2} \right].  
\label{MFSingleLinear}
\end{equation}
We plot the void size distribution obtained with equation
(\ref{MFSingleLinear}) as the thick solid curve in Figure \ref{Ap}.  It is indeed virtually identical to the cases where $A_p =
1.06$ and $1.69$.

The weak dependence of our results on the void-crushing barrier also allows us
to obtain an approximate analytic expression for the fraction of mass
contained within voids with masses greater than $M$. In what follows, we
neglect the void-crushing barrier entirely and assume a single linear void
barrier with $A_v < 0$ and $\beta
> 0$.  Following the appendix of \citet{McQuinn2005}, the fraction of trajectories that cross the void-barrier between scales of $S$ and $S + dS$ is
\begin{equation}
F_v(S) dS = -\frac{d}{dS}\int_{A_v \beta}^{\infty}{\frac{dy}{\beta} Q_{lb}}~dS.
\label{FirstCrossing1LB}
\end{equation}
where 
\begin{eqnarray}
Q_{lb}(y,S) & = & \left\{ \exp\left[ - \frac{y^2}{2 \beta^2 S} 
\right]  \,- \, \exp\left[-
    \frac{(y - 2 A_v \beta)^2}{2 \beta^2 S} \right] \right\} \nonumber
\\ & & \times \frac{1}{\sqrt{2 \pi S}} \exp\left[-\frac{\beta^2 S}{2} - y \right].
\label{Q_lb}
\end{eqnarray}      
Note that our limits of integration differ from those in equation (C11) of
\citet{McQuinn2005} since $A_v < 0$.  Integrating equation
(\ref{FirstCrossing1LB}) from $0$ to $S$ yields
\begin{eqnarray}
F(>M) & = & 1 \, - \, \frac{1}{2} \mathrm{erfc}\left[ \frac{A_v + \beta \sigma^2(M)}{\sqrt{2}~
    \sigma (M)} \right] \,  + \, \frac{\exp(-2 A_v \beta)}{2} \nonumber \\ & &
\times \left( 1 + \mathrm{erf}
  \left[\frac{A_v - \beta \sigma^2(M)}{\sqrt{2}~\sigma (M)} \right] \right).
\label{AnalyticFvol}
\end{eqnarray} 
In most cases of interest, equation (\ref{AnalyticFvol}) quite accurately
approximates the fraction of mass contained within voids
with mass $m > M$ at high redshifts (including the void-crushing barrier).  Note, however, that the fraction of space containing voids larger than a given radius is not as straightforward to compute, because the volume conversion factor $\eta$ is a function of the void mass.
  
\section{Stochastic fluctuations in the galaxy distribution}
\label{StochasticVoids}                        
\subsection{Stochastic voids}
\label{PoissonVoids}

The small galaxy densities in Table \ref{MminTable} reflect the fact that
observable galaxies are increasingly sparse at high redshifts.  Their random
fluctuations will form large empty regions, henceforth referred to
as ``stochastic voids''.  These are inherently different from the voids we
model with the excursion set approach.  They do not form gravitationally and
do not yield any useful information on large-scale structure.  For the range of
galaxy densities we consider, they are a major source of noise that could
potentially obscure meaningful measurements in high-redshift surveys.  In this
section, we estimate the sizes of typical stochastic voids in order to
determine the likelihood that they will be misidentified as real voids.

Our first task is to define a stochastic void properly.  Suppose that galaxies
were truly randomly distributed, obeying Poisson statistics.  In that case,
the probability that a region of comoving volume $V$ and mean galaxy number
density $n_g$ will contain zero galaxies is given by the well known formula
\begin{equation}
P_0(V) = \exp\left[- n_g V \right].
\label{StandardPoisson}
\end{equation}

Equation (\ref{StandardPoisson}) does not account for the fact that an empty
region may lie inside of a larger empty region.  In order to avoid
over-counting smaller stochastic voids (much like the void-in-void problem), we
define a stochastic void as the largest sphere that will fit inside of an empty
region in a random distribution of galaxies.  The probability that a
stochastic void will have a radius between $R$ and $R + dR$ is the probability that a sphere of
radius $R$ is empty multiplied by the probability of encountering at least one
galaxy when the radius of the sphere is enlarged by $dR$.  The latter is
simply $n_g 4 \pi R^2 dR$.  Thus, the probability that a stochastic void will
have a radius between $R$ and $R + dR$ is given by
\begin{equation}
dP(R) = n_g 4 \pi R^2 \exp\left[ - n_g (4 \pi / 3) R^3 \right]~dR.
\label{PoissonVoid}
\end{equation}

\begin{figure}
\begin{center}
\resizebox{8.5cm}{!}{\includegraphics{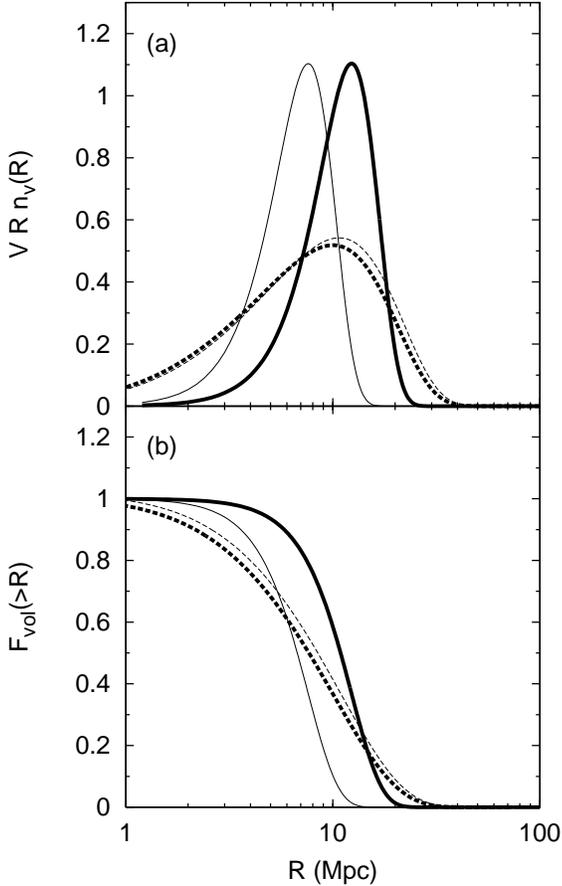}}
\end{center}
\caption{$(a)-(b)$: Size distributions and volume filling fractions of true (dashed) and stochastic (solid)
  voids for $z = 5.7$ (thin) and $z = 10$ (thick).  At higher redshifts,
  stochastic voids are typically the same scale as true voids, making the
  identification of large-scale structure difficult.}
\label{PoissonPlot}
\end{figure}

To obtain a quantity that is directly comparable to our previous calculations,
we consider the stochastic void probability per logarithmic interval in $R$.  Figure \ref{PoissonPlot} $(a)$ shows $dP(R)/d\ln R$ at $z = 5.7$ and $z
= 10$.  The fraction of space filled by stochastic voids larger than $R$,
obtained by integrating equation (\ref{PoissonVoid}) from $R$ to
$\infty$, is shown in panel $(b)$.  For both plots, the solid curves
correspond to stochastic voids while the dashed curves represent true voids.
The thin and thick curves correspond to $z = 5.7$ and $z = 10$ respectively.  At $z = 10$ we assume a halo mass threshold of $m_{min} = 10^{10} \Msol$.

%%%%%%%%%%%% FIGURE 6
%%%%%%%%%%%%%%%%%%%%%

Panel $(a)$ shows that random fluctuations in the galaxy distribution are
slightly smaller than true voids at $z = 5.7$, making the identification of
real voids with radii below the characteristic size difficult.  However, the stochastic void
distribution displays a sharp cutoff at $R \sim 10 \Mpc$ due to exponential suppression.  Thus, in order to minimize the contamination of void samples with stochastic voids, it is
necessary to seek real voids with scales of $R > 10 \Mpc$.  This is, of course, a model-dependent statement:  if we imposed a less rigorous definition for ``true'' voids (allowing them to be, say, only 50\% underdense), they would become larger and more easily differentiable from stochastic voids.

As we discuss in \S \ref{VoidDef}, we assume a lower luminosity
limit of $3\times10^{42}$ ergs/s when calculating the real void size
distribution for $z = 5.7$.  The luminosity limit was chosen to be consistent
with the detection threshold reported by \citet{Shimasaku2006}.  
Lower detection thresholds reduce the characteristic scales of stochastic
voids more than real voids.  While real voids sizes decrease by a maximum of
$\sim 30 \%$ for a lower luminosity limit of $3.75 \times 10^{41}$ ergs/s,
stochastic void sizes decrease by $\sim 50 \%$.  Surveys with lower detection thresholds are therefore much better suited to identify real voids.  

The situation at higher redshifts depends on the halo mass detection
threshold.  At redshifts of $z = 7$, $8$, $9$, and $10$, the characteristic
scales of stochastic voids are roughly equal to those of real voids for
$m_{min} = 6.0 \times 10^{10}$, $3.0 \times 10^{10}$, $1.5 \times 10^{10}$,
and $8.2 \times 10^9 \Msol$ respectively.  For $m_{min}$ larger than these
values, stochastic voids have larger characteristic scales than real voids and
vice versa.  Thus, future surveys must obtain increasingly lower detection
thresholds in order to obtain useful information on void properties at $z = 7
- 10$.   

For completeness, we note that the Sheth-Tormen mass function yields smaller
stochastic void sizes due to the increased mean number density at higher
redshifts (see Table \ref{MminTable}).  The results presented in
Figure \ref{PoissonPlot} are also quite sensitive to the choice of $\sigma_8$
because galaxy densities are extremely sensitive to this parameter at these
redshifts.  

\subsection{Empty voids}
\label{EmptyFrac}

The calculations above suggest that a significant
portion of the high-redshift sky should consist of voids -- both true and
stochastic.  Although we have defined the latter to be completely empty
regions (in contrast to true voids, which just have small overall densities),
in practice high-redshift galaxies are so rare that the two will be difficult
to distinguish.  For example, we expect only 0.45 and 0.11
galaxies inside each void with a diameter of $20 \Mpc$ at $z = 5.7$ and $z=10$
(assuming $m_{min} = 10^{10} \Msol$ for the latter).  The similar appearances of both kinds of voids, together with the relatively small observational samples so far obtained, suggests a modified definition of the void barrier that accounts
for the stochastic nature of the galaxy number counts inside of voids.  

Consider an ensemble of underdense regions with comoving volume $V$ and a mean
comoving galaxy density of $n^v_g = n_g (1+ \dG)$. \citet{CasasMiranda2002}
found that the galaxy number variance in underdense regions at the present day
is very close to the Poissonian value.  We therefore assume that the galaxy
number is Poisson distributed about the mean value $n^v_g V$. The
probability that an underdense region contains zero galaxies is given by
equation (\ref{StandardPoisson}) with $n_g \rightarrow n^v_g$.  Writing
$P_0(V)$ in terms of the mean galaxy underdensity $\dG$ and the total mean
galaxy density $n_g$ yields
\begin{equation}
P_0(V) = \exp\left[- n_g (1 +  \dG) V \right].
\label{EmptyPoisson}
\end{equation}

%%%%%%%%%%%% FIGURE 7
\begin{figure}
\begin{center}
\resizebox{8.5cm}{!}{\includegraphics{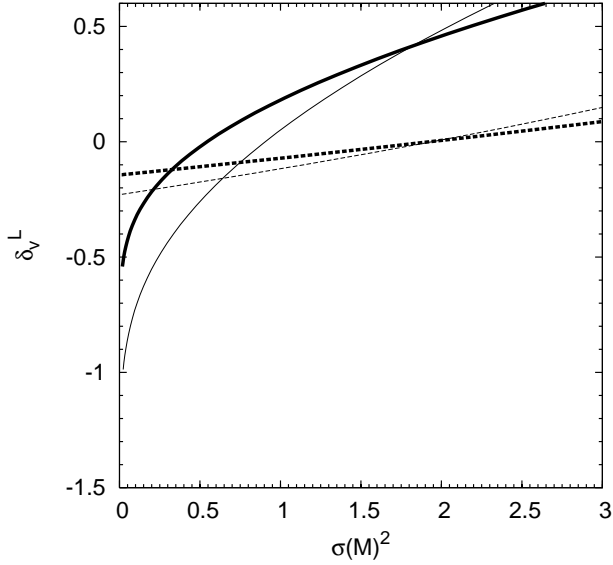}}
\end{center}
\caption{Void barriers (pre-extrapolation) at $z = 5.7$ (thin) and $z = 10$ (thick) assuming that 50\% (solid) of voids at all
  scales are observed to be completely empty.  The dashed lines correspond to the fiducial model with
  $\dG$ = -0.8.}
\label{VoidBarriers_EF}
\end{figure}
%%%%%%%%%%%%%%%%%%

Equation (\ref{EmptyPoisson}) suggests a simple
way to define voids in terms of the probability that a region will be
completely empty.  In our new definition of voids, we fix the value of $\dG$
such that the probability $P_0$ that a region of space will be empty is constant
for all scales.  Inverting equation (\ref{EmptyPoisson}) to find $\dG$ in terms of $R$ and $P_0$ yields
\begin{equation}
\dG(R, P_0) = -\frac{3 \ln P_0}{4 \pi R^3 n_g} - 1.
\label{newdelta_gal}
\end{equation}
For a fixed $P_0$, we plug equation (\ref{newdelta_gal}) into equation
(\ref{deltaL}) to obtain the dark matter density contrast $\delta^L_v$ required
to produce an average galaxy underdensity of $\dG(R, P_0)$.  We then fit
$\delta^L_v$ to a linear function of $\sigma^2$ such that $\delta^L_v$ is well
approximated in the $R \approx 5 - 20 \Mpc$ regime.

The void distributions we obtain using equation (\ref{newdelta_gal}) actually contain both stochastic and real voids.  The former and latter dominate the distribution at small and
large $R$ respectively.  The transition between the two regimes occurs at the
scale for which $\dG(R) \sim 0$, or $R_0 \sim \left[3 \ln (1/P_0)~/~4 \pi
  n_g\right]^{1/3}$.  For $R > R_0$, $\dG$ quickly approaches $ -1$. Voids with scales such that $\dG(R) < -0.8$
represent real galaxy voids as we have defined them in \S \ref{VoidDef}. For $R <
R_0$, the mean galaxy underdensity is greater than zero.  Although regions with this scale have a probability $P_0$
of being empty, they are, on average, overdense in both galaxies and dark matter.  Hence, the voids
obtained in our new model with scales $R < R_0$ are statistical fluctuations
that result from finite size effects.  We note that equation
(\ref{newdelta_gal}) is not a perfect definition of empty voids since a
fraction $1- P_0$ of voids will contain galaxies.  Nonetheless, it is the
closest we can come to defining voids as empty regions in the excursion set formalism. 

%%%%%%%%%%%%%% FIGURE 8
\begin{figure}
\begin{center}
\resizebox{8.5cm}{!}{\includegraphics{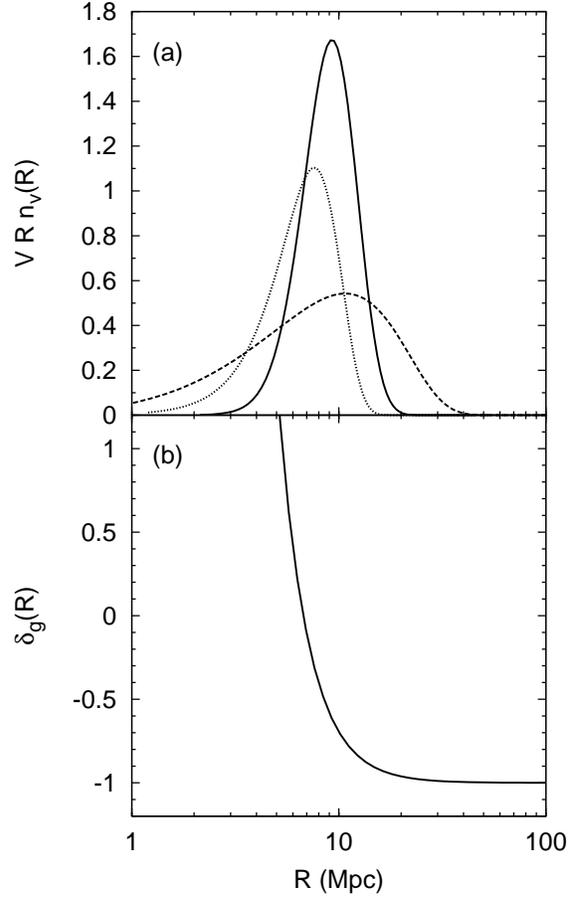}}
\end{center}
\caption{$(a)$: Void size distributions at $z = 5.7$.  The solid, dashed, and
  dotted curves correspond to modified, fiducial, and stochastic void distributions
  respectively. $(b)$: The average galaxy underdensity such that voids of all
  scales have a 50\% probability of being empty.}
  \ref{VoidBarriers_EF}
\label{VoidDistributions_EF}
\end{figure}
%%%%%%%%%%%%%%%%%%%%%%%%

The thin and thick solid curves in Figure \ref{VoidBarriers_EF}  show the void
barriers at $z = 5.7$ and $z = 10$ respectively in our new definition.  We use
the parameters from Table \ref{MminTable} and an empty fraction of $P_0 =
0.5$.  The dashed curves in Figure \ref{VoidBarriers_EF} show the fiducial model
void barriers.

The solid and dashed curves in Figure \ref{VoidDistributions_EF} $(a)$ show the void
distributions derived from the new and fiducial void barriers at $z = 5.7$
respectively.  The dotted curve shows the stochastic void distribution
obtained in \S \ref{PoissonVoids}.  For reference, we show the mean galaxy
underdensity required by equation (\ref{newdelta_gal}) in panel $(b)$.  As Figure \ref{VoidDistributions_EF} shows, the void distribution obtained
with equation (\ref{newdelta_gal}) has a characteristic scale that is slightly
larger than the stochastic distribution.  Both panel $(b)$ and the dotted curve in panel $(a)$ illustrate that the newly obtained distribution is dominated
by stochastic fluctuations for $R \la 7 \Mpc$.  The newly calculated void
distribution is also more sharply peaked than the fiducial curve.  The steeper
large scale cutoff is due to the fact that, in our new definition, larger voids
must be more underdense in order to have a 50 \% probability of being empty.
Such underdense regions are increasingly rare at large $R$.  We emphasize,
however, that larger voids do exist as predicted by the fiducial model; they
just do not satisfy our new criteria.    

The purpose of this section has been to show that stochastic voids present a significant contaminant at high redshifts.  The expected sizes of deep large-scale voids (at least 80\% underdense in galaxies) are always fairly close to the sizes of empty regions.  Searches for large-scale structure must therefore either (1) confine themselves to extremely large scales ($\gg 10 \Mpc$) and modest underdensities or (2) reach sufficient depth to detect small halos ($\sim 10^9 \Msol$).  The latter may in fact be possible if the lensed sources observed by \citet{Stark2007} prove to be at $z=9$, suggesting number densities $\sim 0.1 \Mpc^{-3}$ \citep{mesinger07}.  

\section{LAEs and the full galaxy population}
\label{VisibleFrac}

In sections \ref{VoidDef} - \ref{VoidDistributions}, we have relied upon high-redshift LAE surveys to
obtain the comoving galaxy number densities for redshifts of $z = 4.86$, $5.7$,
and $6.5$.  Thus far we have assumed that all galaxies within haloes
with masses greater than the threshold $m_{min}$ are observable in these surveys.  However,
lower redshift studies suggest that only a fraction of galaxies have
strong Ly $\alpha$ emission lines.  For example, \citet{Gawiser2007} estimate that only $\sim 1 - 10 \%$
of haloes with masses above $\sim 10^{10.6}\Msol$ are occupied by LAEs at $z = 3.1$.  It is therefore
reasonable to suspect that many high-redshift galaxies go undetected in LAE
surveys.  In this section, we consider what happens to the void size
distributions when we assume that only a fraction of galaxies are sampled in
LAE surveys.  We will assume throughout that the processes determining whether or not a particular halo hosts an LAE are internal to the galaxy itself and so are independent of its environment.

%%%%%%%%%%%% FIGURE 9
\begin{figure}
\begin{center}
\resizebox{8.5cm}{!}{\includegraphics{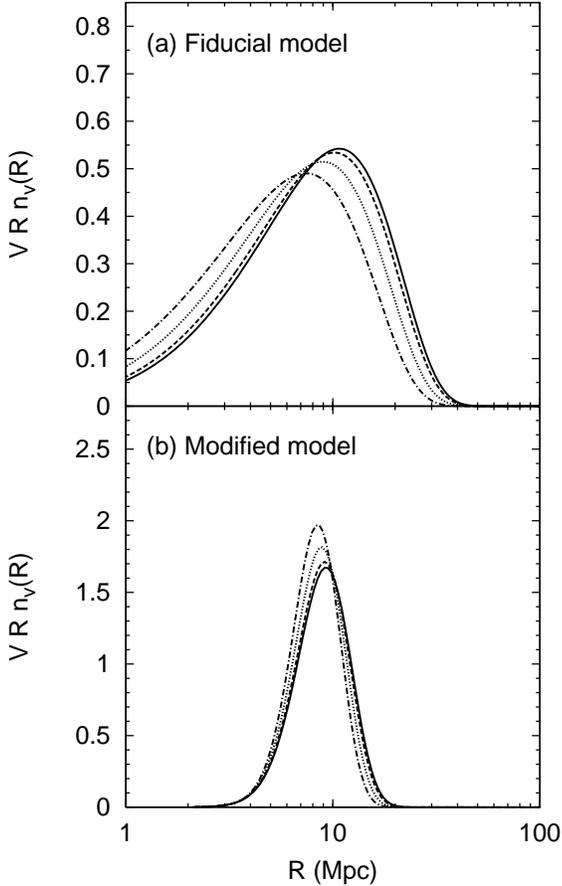}}
\end{center}
\caption{Void size distributions in LAE surveys at $z = 5.7$ assuming that
  only a fraction of galaxies are sampled.  The solid, dashed, dotted, and
  dot-dashed models assume that 100 \%, 70\%, 30\%, and 10\% of galaxies are
  LAEs respectively.  The characteristic scales of true voids decrease if only
a fraction of galaxies are detected, increasing the contamination of void samples by stochastic fluctuations.}
\label{fvisPlot}
\end{figure}
%%%%%%%%%%%%%%%%%%%%%%

In what follows, we assume that a randomly chosen fraction of galaxies are
LAEs.  Mathematically, this decreases the mean observable galaxy density derived from LAE surveys by a factor $f_{vis}$, from now on referred to as the visible fraction.  The observable mean galaxy density is given by
\begin{equation}
n^o_g(m_{min}) = f_{vis}~n_g(m_{min}),
\label{fvis}
\end{equation}
where $n_g$  is the intrinsic mean galaxy density and the superscript ``o''
denotes observable.  The mean galaxy density within an
underdense region $n_g(m_{min}|\delta^L,M)$ decreases by the same factor $f_{vis}$. 

The procedure for calculating void size distributions is only slightly
modified from the usual case, because the halo mass
threshold is now a function of $f_{vis}$.  For a fixed observable mean galaxy
density and visible fraction, equation (\ref{fvis}) is solved for
$m_{min}(f_{vis})$.  As before, we define voids in terms of the
\emph{observable} mean galaxy underdensity through equation (\ref{deltaL}).   

Figure \ref{fvisPlot} $(a)$ shows the void
size distributions at $z = 5.7$ for a variety of visible fractions
in the fiducial model.  Panel $(b)$ shows the modified model of \S
\ref{EmptyFrac}.  Both panels in Figure \ref{fvisPlot} show that the visible
fraction has only a small effect on the void size distributions.  Even under
the assumption that 10\% of LAEs emit, there is only a $\sim 30\%$ difference
in characteristic scales with the fiducial curves, and even less with the modified model.  The visible fraction does however have an
important effect on the role of stochastic voids.  Since we have held the
observable galaxy number density fixed, the corresponding stochastic void
distributions are unaffected by $f_{vis}$.  Therefore, as $f_{vis}$ decreases
and the characteristic scales of real voids gets smaller, void samples are increasingly contaminated by stochastic fluctuations.    

\section{Comparison to Observations}
\label{Observations}

Owing to a lack of statistics, a rigorous comparison of our calculations
to observational surveys is not possible.  In this section, we content ourselves with a rough comparison to the surveys of \citet{Shimasaku2006} and \citet{Ouchi2005}.  We focus on $z = 5.7$ due to a number of recent claims of large-scale structure at this redshift.    

\citet{Shimasaku2006} report evidence for the existence of large scale structure at
$z = 5.7$ in their photometric sample of 89 LAEs, including 34
spectroscopically confirmed objects.  Their argument is based on a roughly 20
\% overdensity and underdensity in the western and eastern halves of their sky
distribution respectively.  Their survey covers a continuous area of 725
arcmin$^2$ and redshifts of $z \approx 5.65 - 5.75$, corresponding to a survey
volume of $1.8 \times 10^5 \Mpc^3$.  We find a volume filling fraction $F(>R)
= 0.76$ for voids with at least half of their survey volume and $\dG = -0.2$.
The large volume filling fraction suggests that our result is consistent with the possibility that the observed underdense region in \citet{Shimasaku2006} is a void progenitor.

\citet{Ouchi2005} report much more well-defined large-scale structure.  Their
catalog of 515 LAEs at $z = 5.7$, which covers an area of 180 Mpc $\times$ 180 Mpc $\times$ 48 Mpc, exhibits a high degree of clustering.   They find
clearly defined voids and filamentary features.  Moreover, the voids depicted
in their survey are extremely large, ranging in size from 10 - 40 comoving Mpc
in scale. Taking the scales of stochastic fluctuations and survey depth into
account, these scales roughly correspond to void volumes of $1.5 - 6.0
\times 10^4 \Mpc^3$, where we have approximated them to be cylindrical regions
with lengths of $48 \Mpc$.  A direct comparison of these voids to our
predictions is problematic since our model assumes a spherical geometry.  The
best we can do is compare void volumes.  The modified distribution shown in Figure \ref{fvisPlot} indicates that the voids of
interest have a radii ranging from $R \sim 10 - 15 \Mpc$, or comoving volumes
of $0.42-1.4 \times 10^4 \Mpc^3$.  Although our modified definition yields
volumes that are slightly smaller than the observed voids, the fiducial
model in Figure \ref{OBS} predicts the existence of a small number of larger
scale voids with comoving diameters and volumes as high as $50 \Mpc$ and $6.5
\times 10^4 \Mpc^3$ respectively.  Thus, we do not consider the large voids
observed by \citet{Ouchi2005} to be in contradiction with our results.  Note as well that this model under-predicts the sizes of $z=0$ voids by a comparable amount \citep{FandP2006}; the discrepancy may be due to redshift space distortions, the non-spherical regions relevant to this narrow-band survey, or our simplified void identification algorithm.

Interestingly, the widest and most recent LAE survey conducted by
\citet{Murayama2007} does not find convincing evidence for the clustering
observed by \citet{Ouchi2005}.  Their survey consists of 119 LAE candidates in
a 1.95 deg$^2$ area, corresponding to a number density of $6.6 \times 10^{-5}
\Mpc^3$.  With such a small sky density (eight times smaller than \citealt{Ouchi2005}), true voids are masked by stochastic fluctuations (which have characteristic scales $\sim 15 \Mpc$).

Finally, we emphasize the difficulty in drawing conclusions from comparisons
to sky distribution maps.   Given the poor statistics, it is often difficult
to determine conclusively whether a given empty region is a real void, and
without redshifts we must compare our (spherical) predicted voids to
cylindrical survey volumes.  Furthermore, the samples we have described here
are not completely spectroscopically confirmed and probably contain a
reasonable fraction of low-redshift contaminants.  When voids are defined
based on only a few galaxies, such contamination can significantly affect the
statistics (and, because the contaminants are also line-emitting galaxies at
discrete redshifts, can introduce their own large-scale structure).  At the
very least, they affect the mass threshold of the survey (although probably
not as much as uncertainty in $f_{vis}$).  Detailed comparisons will require
simulations of the effects of these contaminants.  The best we can say now is that there is no inconsistency with our model.  Future surveys will
undoubtedly allow for a more systematic comparison.

\section{Discussion}
\label{Discussion}

We have calculated void size distributions at $z = 4.86$--$10$ using the excursion set model developed by \citet{SandV2004} and
\citet{FandP2006}.  The latter found characteristic void radii of $R \approx  7-14 \Mpc$ at $z = 0$.  For the
observational sensitivities assumed in this paper, we obtained
characteristic void radii that are very similar:  $R \approx 7-10 \Mpc$ for
redshifts between $z = 4.86$ and $z = 10$.  These results are virtually
independent of the void-crushing barrier (for any reasonable choice).  We have
shown that characteristic void scales actually increase with redshift for a
fixed halo mass threshold due to a decreased number density and increased bias
with respect to the underlying matter density.  Following recent studies on
the abundances of low-redshift LAEs, we explored the possibility that only a
fraction $f_{vis}$ of galaxies are sampled in LAE surveys.  This has only a small effect on the void size distribution but increases the contamination of void samples by stochastic fluctuations.  

In section \ref{StochasticVoids}, we have explored stochastic fluctuations in
the galaxy distribution.  These fluctuations, although inherently different
from the "real" voids we model in this paper, will result in large empty
regions in the sky. Stochastic voids can therefore contaminate real void
samples and lead to erroneous conclusions on the formation of large-scale
structure.  We have estimated the typical scale of these regions to be
slightly smaller than the characteristic scale of true voids at $z \sim 5$.
At $z \sim 10$, the situation depends on the particular choice of $m_{min}$.  For $m_{min} \sim 10^{10} \Msol$, stochastic voids are typically the
same scale as real voids.  The increased importance of
stochastic fluctuations will make the identification of large-scale structure
at this redshift difficult.  Attempts to do so must observe halos near the minimum mass to form stars, $\sim 10^8$--$10^9 \Msol$, in order for true voids to dominate the observed distribution.  

We found that a large fraction of real voids in our fiducial model contain no
visible galaxies, adding to the difficulties in differentiating them from
stochastic fluctuations.  We have presented a modified definition of voids
that incorporates both stochastic and real voids and so is easier to compare to the limited observational samples thus far available.  In our new approach, we defined voids in terms of the probability for a region to be empty.  We found
that the modified void distributions are more sharply peaked and have
characteristic scales that are comparable to the fiducial model.    

We have also attempted to visually compare our results to the most recent
narrow-band filter surveys at $z = 5.7$.  While we found no inconsistencies,
it is difficult to draw any decisive conclusions because of small-number
statistics, projection effects, and lower-redshift contaminants.  Obviously, a
more systematic approach is required.  Future surveys promise to provide
better statistics and increased sample volumes for studies on high-redshift voids.

In the context of next generation surveys for high-redshift galaxies, our model is useful for gauging the impact of cosmic variance.  Consider a fictitious survey at $z = 10$ with a detection
threshold of $m_{min} = 10^{10} \Msol$.  Figure \ref{PoissonPlot}
illustrates that stochastic voids with $R \sim 10 - 20 \Mpc$ will dominate the
sky distribution.  Therefore, one must either search for voids with $R > 20
\Mpc$ or search deeper for significantly smaller sources.  The latter may be
possible if the sources observed by \citet{Stark2007} are indeed at $z \sim
9$, in which case they imply that halos near $\la 10^9 \Msol$ are visible
\citep{mesinger07}.  However, high-redshift galaxies are so highly biased that
even with deep observations, a substantial fraction of the Universe is filled
with empty or nearly-empty regions.  For example, at $z=10$ and $m_{min} =
10^{10} \Msol$, $\sim 37\%$ of space is filled by regions that are at least
80\% underdense in galaxies and at least 20 Mpc across -- or fully 7 arcmin.  With the small fields of view available to near-infrared detectors, this suggests that either many independent fields must be observed or a large contiguous volume surveyed to be guaranteed of detecting a reasonable number of sources.

Finally, we have neglected reionization and its
effect on the appearance of large-scale structure.  Regions of neutral
hydrogen are expected to modulate the LAE density on large scales and
accentuate the appearance of structure
\citep{furl04-lya, Furlanetto2006,McQuinn2006,McQuinn2007, mesinger07}.
Although the precise time frame is currently unknown, quasar observations and
cosmic microwave background measurements have provided some evidence that
reionization occurred between $z \sim 6 - 10$ (e.g,
\citealt{Fan2006,Page2007,MandH2004,MandH2007}).  Interestingly,
\citet{Kashikawa2006} found a significant high-luminosity suppression in the LAE
luminosity function between $z = 5.7$ and $z = 6.5$. Whether or not
reioinization is responsible for this effect is currently unclear \citep[no such suppression was observed by][]{dawson07}.

Because IGM absorption modulates the LAE density on large scales, we would
expect reionization to have a substantial effect on the observed void sizes in
such narrow-band surveys (it should not affect galaxies identified through
broadband effects).  Of course, the plots in Figure \ref{Z7to10} provide
analytic estimates only of the \emph{intrinsic} void size distributions.  They
provide a basis for comparison with high-redshift surveys in order to
determine whether the observed features are easily attributable to the
large-scale clustering alone.  It therefore helps illuminate efforts to use
voids to constrain the IGM properties during reionization, as first attempted
by \citet{McQuinn2007}.

\bibliographystyle{mn2e}
\bibliography{voids}

\end{document}